# Estimate of the impact of background particles on the X-Ray Microcalorimeter Spectrometer on IXO

S. Lotti [a], E. Perinati [b], L. Natalucci [a], L. Piro [a,c], T. Mineo [d], L. Colasanti [a], C. Macculi [a] *

[a] *INAF-IAPS Roma, Via fosso del cavaliere 100, Rome, 00133, Italy*

[b] *IAAT - Institut für Astronomie und Astrophysik, Universität Tübingen, 72076 Tübingen, Germany*

[c] *Astronomy Department, Faculty of Science, King Abdulaziz University, P.O. Box 80203, Jeddah 21589, Saudi Arabia*

[d] *INAF-IASF Palermo, Via Ugo la Malfa 153, Palermo, 90146, Italy*

---

**Abstract**

We present the results of a study on the impact of particles of galactic (GCR) and solar origin for the X-ray Microcalorimeter Spectrometer (XMS) aboard an astronomical satellite flying in an orbit at the second Lagrangian point (L2). The detailed configuration presented in this paper is the one adopted for the International X-Ray Observatory (IXO) study, however the derived estimates can be considered a conservative limit for ATHENA, that is the IXO redefined mission proposed to ESA. This work is aimed at the estimate of the residual background level expected on the focal plane detector during the mission lifetime, a crucial information in the development of any instrumental configuration that optimizes the XMS scientific performances. We used the Geant4 toolkit, a Monte Carlo based simulator, to investigate the rejection efficiency of the anticoincidence system and assess the residual background on the detector. 

---

* Simone Lotti. Tel.: +39-06-49934690; e-mail: simone.lotti@iasf-roma.inaf.it

## 1. Introduction

The X-Ray Microcalorimeter Spectrometer [1] was one of the focal plane instruments planned for the IXO mission. The design of the detector is based on an inner and an outer Transition Edge Sensor (TES) array, lodged in two different levels. The inner array is composed of 40 x 40 pixels, with a pixel size of 300 μm, covering 2 x 2 $arcmin^2$ Field of View (FoV), while the outer array surrounds the inner array as a frame with a 52 pixels side, each of 600 μm, enlarging the FoV to 5x5 $arcmin^2$. The physical size of the array is 31.2 x 31.2 $mm^2$. Each pixel in the array is a microcalorimeter made of a TES in thermal contact with a 7 μm thick absorber (6 μm Bi and 1 μm Au). XMS aims to perform high resolution spectroscopy (with an energy resolution of 2.5 eV) in the energy band 0.2-10 keV.

TES detectors do not distinguish among different particles and photons that release energy inside the detector band pass, and therefore to reject fake signals produced by cosmic particles and enhance the S/N ratio a new concept Anti-Coincidence Detector (ACD) was foreseen, composed of a TES array of 4 large area pixels for a total surface of a 36 x 36 $mm^2$, placed 2 mm below the detector. This way particles cross through the detector and the ACD (unlike photons that are completely absorbed inside the main detector), causing a simultaneous signal that allows to discriminate the event.

No X-ray missions were flown to L2 up till now, so in order to investigate the anticoincidence rejection efficiency, to predict the level and identify the major sources of unrejected background, we built a Monte Carlo simulator based on the 9.4 version of the Geant4 software and performed several sets of simulations. The physics was validated through simple tests of the physics involved (see Section 2) and reproducing the background measured by the XRS microcalorimeter flown on Suzaku to within 25% in total flux in a qualitatively similar spectrum [2].

In April 2012 IXO has undergone a redefinition phase under the new name of ATHENA [3]. With respect to IXO the configuration of XMS is different in various aspects. Those most directly affecting background performances are essentially two. The area of the ATHENA TES array detector will be a factor of 6 times smaller. In addition, there will be only a single TES array, rather than two. Therefore the distance between the anticoincidence and the array will be a factor of 2 smaller than assumed here with reference to the IXO configuration. In addition since the ATHENA will have roughly half the focal length of IXO (11.5 m instead of 20 m) an $arcsec^2$ of sky covers more detector area on IXO than ATHENA, and thus the same background in terms of $cts/cm^2/s$ will result lower on ATHENA in terms of $cts/cm^2/s/arcsec^2$. In this respect estimates derived here for IXO can be considered as a conservative limit when applied to ATHENA. Furthermore, this work already allows to identify possible modifications aimed to reduce the unrejected background with the new configuration of the detector. A specific study tailored to the ATHENA configuration will be subject of a future paper.

## 2. Validation of Geant4 physical models involved

TES detectors are sensitive to any energy released inside the absorber, that includes not only the photons we want to detect, but also signals from fluorescence photons produced in the surrounding material or charged particles passing through the detector. These can be primary particles passing through the spacecraft and reaching the focal plane, or secondaries created by the interaction of primaries onto the materials surrounding the detector. The first have their energy degraded by interactions with the spacecraft material before releasing their energy - mostly by ionization - into the absorber. The second are composed mainly of electrons and fluorescence photons. These secondary particles have energies much lower than the corresponding primaries and usually are completely absorbed inside the detector without reaching the anticoincidence.

Therefore, in order to obtain realistic simulations, it is mandatory to have a correct treatment of electromagnetic interactions, especially at low energies, and of fluorescence production. There are two Geant4 models that allow to simulate the low energy processes: Livermore and Penelope, and both provide precise treatment of electromagnetic interactions at low energies[1]. We tested both of them and found negligible differences (less than 0.5%) in the detector output, so in the simulations we decided used the Livermore package since it provides wider energy extension.

———

[1] For a detailed explanation of the packages differences refer to https://twiki.cern.ch/twiki/bin/view/Geant4/LowEnergyElectromagneticPhysicsWorkingGroup

Table 1. Simulated and experimental ratios of fluorescence line photons produced $N_\gamma$ over the number of impacting particles $N_i$.

| Impacting particles | $N_\gamma/N_i$ (Experimental) | $N_\gamma/N_i$ (Geant4) |
|---|---|---|
| $e^-$ (45 keV) on Cr | $64.6 \pm 3.2 \times 10^{-4}$ | $58.2 \pm 2.4 \times 10^{-4}$ |
| Protons (3 MeV) on Cu | $67.1 \pm 2.6 \times 10^{-4}$ | $55.3 \pm 2.4 \times 10^{-4}$ |

In order to verify the correct behavior of Geant4 in the treatment of fluorescence photons production from electrons and protons (PIXE) we run a series of simple tests, and compared the results with experimental data found in literature. In order to reproduce experimental results reported in [4] we simulated a Cr slab impacted by 45 keV electrons, and a Cu slab hit by 3 MeV protons [5]. We report in Table 1 simulated and experimental ratios of detected fluorescence photons per impacting particle. The differences, of the order of 10-20%, are due to uncertainties in the experimental setup used in the references and to practical difficulties in separating line photons from the continuum.

Single event interactions, such as those produced by interaction of neutrons with the absorber of the TES array, are potential sources of residual background because these events cannot produce a simultaneous signal in the anticoincidence. Elastic scattering of neutrons gives a recoil of the nucleus that can produce a signal in cryogenic detectors [6]. The interaction of highly energetic particles with the materials surrounding the detector can lead to generation of neutrons that can impact the detector. We have therefore introduced the treatment of hadronic interactions in the simulations. The models used to handle the hadronic interactions are the Bertini model at low energies (0 - 9.9 GeV), the low energy parameterized model at intermediate energies (9.5 - 25 GeV) and the Quark-Gluon String Model at higher energies (15 GeV – 100 TeV).

We also performed simple tests on neutron production yields and neutron cross section. Neutron production from protons has been simulated on Tungsten at several energies, recreating the experimental setup similar to the one reported in [7]. The results obtained showed the same trend reported by the authors, with numerical differences inside 8%. Neutron interactions were tested simulating a Nb slab impacted by 30 MeV neutrons, and the results from the simulation agreed with theoretical calculations within a 8%. The differences are attributable to the fact that we considered the cross section at a fixed energy, while the particles lose energy travelling in the material. [2]

## 3. Particle environment in L2

The mission will be flown in L2 orbit, where the radiation environment presents two main contributions [8]. The first component of radiation background in L2 are cosmic particles with galactic origin (GCR). Most of them are hard protons, α-particles and electrons, with energies in the range 10 MeV - 100 GeV. These particles have enough energy to penetrate the spacecraft walls and the cryostat, creating secondary particles along their way.

Table 2. Expected integral fluxes of particles in the L2 orbit

| Particle | Flux (p/cm$^2$/s) 60 MeV – 100 GeV |
|---|---|
| Protons | 4.2 |
| α-particles | 0.092 |
| Electrons | 0.15 |

The other component, soft solar protons, is variable depending on the intensity of solar activity. During flares the flux of soft protons can increase by several orders of magnitude with respect to the average value. It is difficult to estimate the flares contribution to the total averaged flux, due to their extremely high variability, occurrence rate and flux. However, most prominent flares occur only for ~ 4 % of the time (according to data from Planck Solar Radiation Environment Monitor) during solar minima and up to ~ 30 % of the time during solar maxima. Since the detector will not be operative during solar flares, we used the solar-quiet proton and α-particles differential fluxes reported by the CREME96 toolkit [9, 10, 11] for the time of IXO/ATHENA launch in 2022, comprehensive of solar and cosmic ray particles. Because 2022 is close to the solar minimum, expected on 2020-1, that corresponds to cosmic rays maximum, and the solar maximum is predicted on 2025-6, the adopted value thus represent a conservative estimate for the mission lifetime. Since CREME96 does not reproduce the flux of cosmic ray electrons, we used the values reported in [12] that describe an averaged flux over an entire solar cycle. All the components are represented in figure 1.

---

[2] To find an accurate description of the models used for hadronic interactions, validation plots and neutrons cross sections visit respectively http://geant4.web.cern.ch/geant4/UserDocumentation/UsersGuides/PhysicsReferenceManual/fo/PhysicsReferenceManual.pdf, http://geant4.web.cern.ch/geant4/results/validation_plots.htm and http://www.nndc.bnl.gov/sigma/index.jsp?as=93&lib=endfb7.1&nsub=10

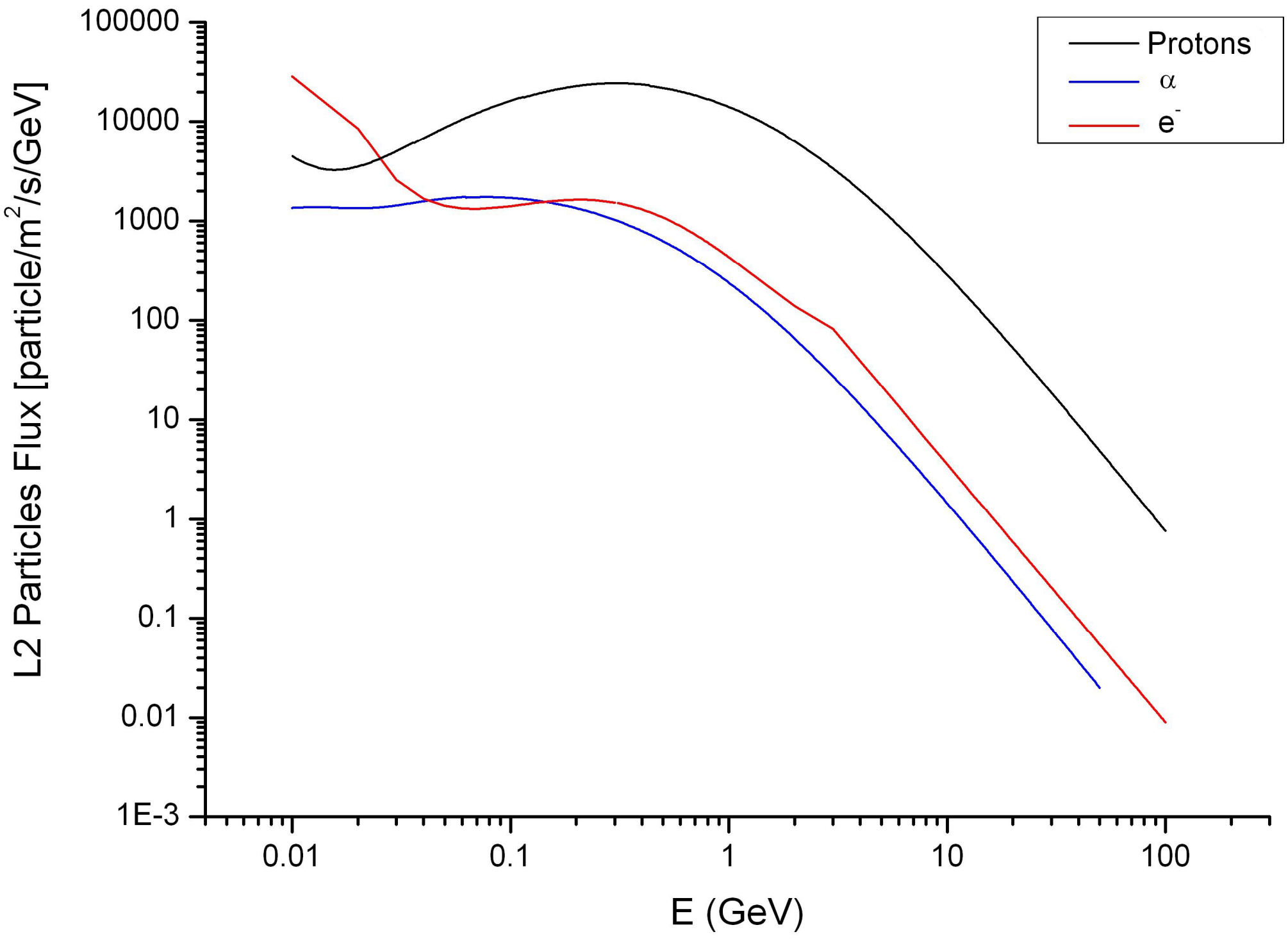

*Fig. 1. Spectra of L2 particles (cosmic rays and solar)*

We integrated the differential fluxes in the simulated energy range and obtained the expected fluxes for the different particles, reported in Table 2. Since the minimum energies required to reach the focal plane is 60 MeV (see Section 4) we used this value as minimum energy for particles. Above this energy the particle spectrum is dominated by GCR.

## 4. The geometrical models

We performed our work with two different approaches: we used a simple geometrical model to represent the structure of the cryostat and of the payload, to run the simulator in relatively small times. We then build a second more detailed model to reach a greater degree of accuracy in our evaluation, but at the cost of an higher computational time. The comparison between the two geometrical models allowed also to evaluate how accurate we needed to model the structure of the cryostat, and the influence of specific changes in the system geometry on the background experienced by the detector.

In the first model – sphere model – we assumed a geometry (fig. 2) composed by 4 concentric spheres. The outer (sphere 1) is 15 mm Al thick, is representing the satellite thickness. Inside that another Al sphere (sphere 2), 1 mm thick, represents the cryostat. The sphere 2 is coated on both sides by two 5 μm thick Nb spheres (sphere 3 and 4). The inner shell of Nb accounts for the innermost shield of the cryostat.

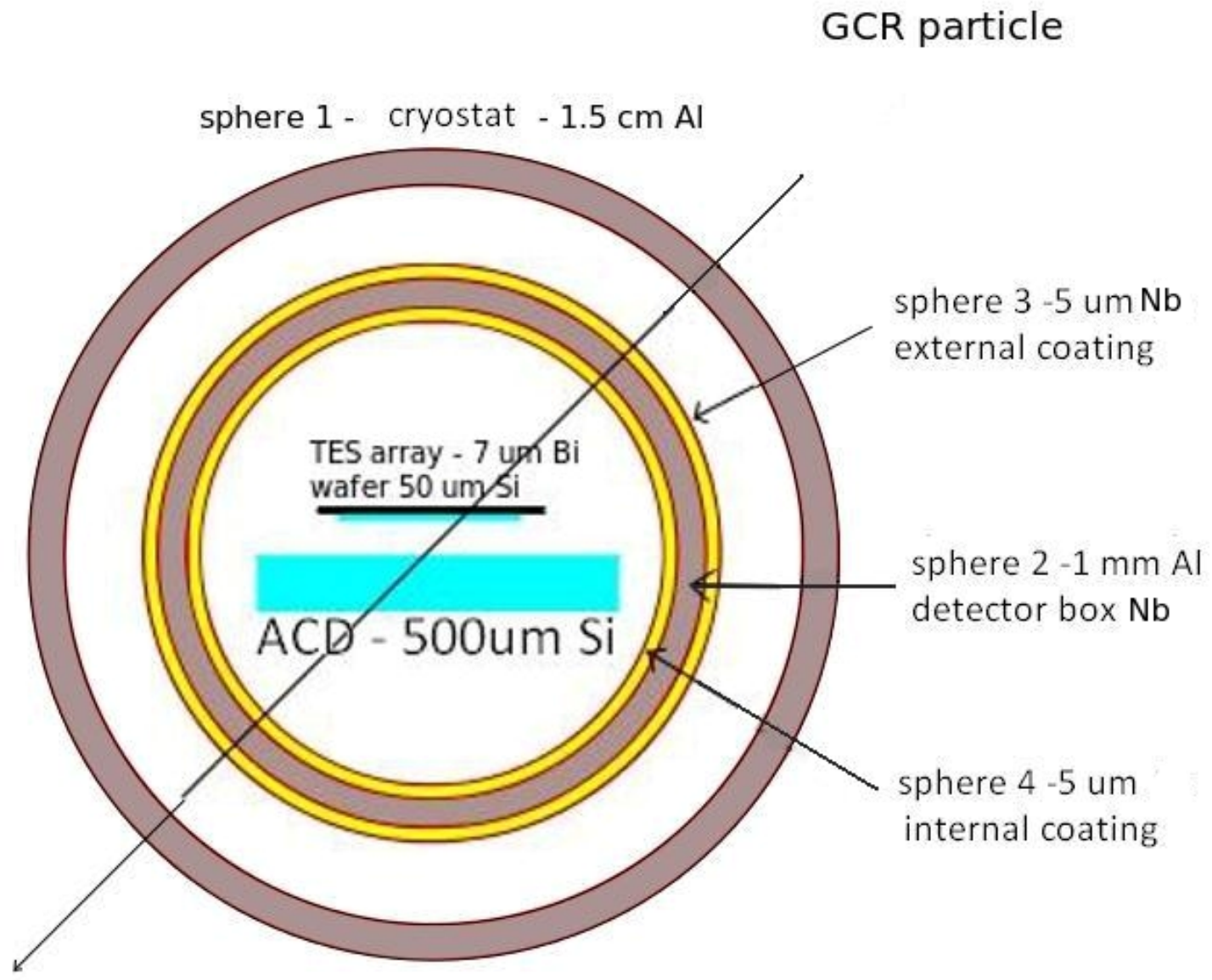

*Fig. 2. The sphere model, representing the spacecraft and the cryostat containing XMS and ACD detectors*

In the second model (fig 3) – detailed model – the cryostat is modeled in great detail (Den-Herder J.W., private communication & Henk van Weers, Focal Plane Assembly trade-off report), while the satellite is still modeled as an Al sphere. For a focal length l = 20 m and a mirror diameter D = 2 m the optics fill a solid angle $\Omega = 0.0312$ sr, a small fraction of the total solid angle, and due to the practical difficulties in their modeling and the low gain in precision we decided to not include them in the model. This is an acceptable approximation since secondary particles have lower energy than the primaries, and the ones generated far from the detector are absorbed by the materials interposed. Therefore an accurate description of the outer structures slows down the simulation with no appreciable benefits [13]. Indeed as we will see later, only a small fraction of the background comes from the sphere representing the spacecraft.

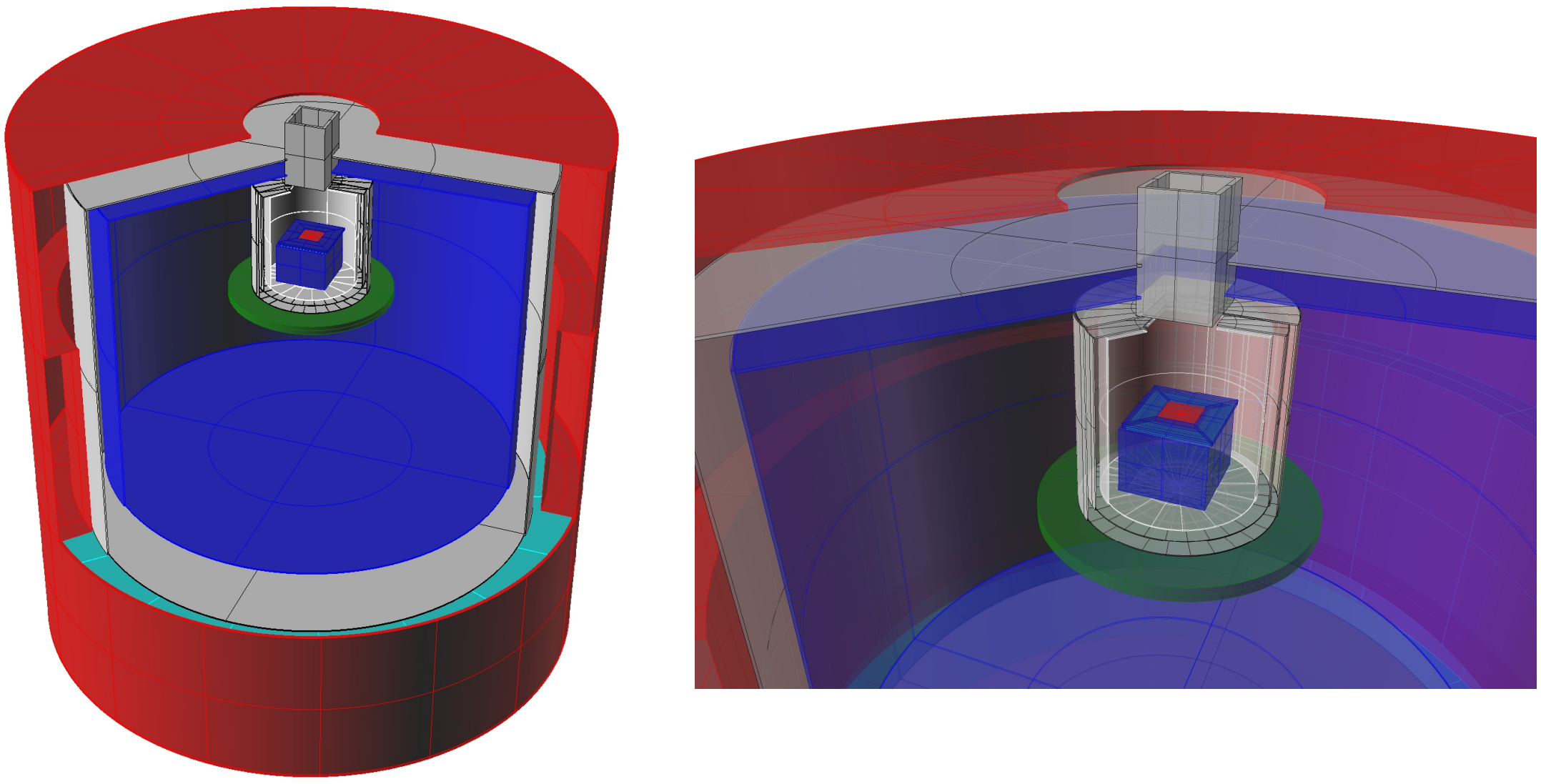

*Fig. 3. The Geant4 detailed model of the cryostat.*
*Left - Three Al external thermal shields are implemented (red, gray and blue).*
*Right – the innermost part of the cryostat: the last stage cooler plate (green), the internal shielding composed of three grey layers (Cryoperm-Aluminum-Niobium from outside to inside). Inside the last shielding the detectors (red), and their supports (blue).*

The two models have different thickness and materials, and provide different cuts in energy for the impacting particles. In the sphere model the thickness of materials encountered by particles directed toward the detector is ~ 1.6 cm, blocking protons below 60 MeV. The thickness encountered in the detailed model however is strongly dependent on the trajectory of the particles, so we reported the mean value that is ~ 150 MeV.

According to XMM-Newton and Chandra experience lower energy solar protons can reach the focal plane being focused by the optics. Their energies remain almost unchanged and the flux is driven by the optics effective area. In this case, to compute the proton rate at the focal plane, a ray-tracing simulator that includes the optics geometry, an appropriate formula for the reflection of the protons from the mirror surface and some evaluation of the scattering angles expected after the reflection is used. The ray-tracing simulations report a focused proton background of 0.01 cts/cm$^2$/s [14].

## 5. Simulations results

We reproduced all the components of the particle environment shown in Section 3 in the simulator, generating a random distribution of incoming directions from a sphere surrounding the entire geometrical model, and processed the detectors output.

Table 3. Particle fluxes experienced in the detector neighborhoods without the anticoincidence detector

| | **Total [cts/cm$^2$/s]** | Primaries [cts/cm$^2$/s] | Secondaries [cts/cm$^2$/s] |
|---|---|---|---|
| Total background on TES array | **5.6** | 4.3 | 1.3 |
| Total background on TES array [0.2-10 keV] | **3.7** | 3.0 | 0.7 |
| Background after autorejection | **4.6** | 3.7 | 0.9 |
| Background after autorejection [0.2-10 keV] | **3.1** | 2.6 | 0.4 |

In the analysis we took into account the TES autorejection besides the ACD rejection efficiency. By autorejection we mean that events causing simultaneous detections in more than one pixel can be rejected. The autorejection is effective on those particles with very skew trajectories that release energy - above the low energy threshold of the TES array, set here at 0.2 keV - in more than one pixel and intercept the TES array but not the ACD.

We first run a simulation on the sphere model to have an estimate of the background level the detector would experience without ACD. We found out that a smart data processing (autorejection) can reduce the background by about 20%. As can be seen from Table 3, the total unrejected background level in the instrument energy range is 3.1 cts/cm$^2$/s, i.e. a factor of 15 above the scientific requirement of 0.2 cts/cm$^2$/s. This result confirms the need for a background rejection tool such as the anticoincidence detector. This value will be taken as a baseline to calculate the rejection efficiency after the insertion of the anticoincidence detector.

### *5.1. Sphere model results*

We run, at first, three different sets of simulations with different distances between TES and ACD on the sphere model in order to show the influence of the distance on the rejection efficiency: $d$ = 1 mm, $d$ = 2 mm, and $d$ = 2.6 mm.

Table 3. Unrejected background induced from the different primary particles: sphere model

| Primaries | Rate (cts cm$^{-2}$ s$^{-1}$) *d* = 2 mm |
|---|---|
| Protons | 0.18 |
| Electrons | 6.3 x $10^{-4}$ |
| Alpha | 9.7 x $10^{-3}$ |
| **Total** | 0.19 |

Considering the ACD veto (assuming an Energy threshold of 20 keV[3] [1]) together with the autorejection and the energy bandpass discrimination of the TES array, we obtained a residual background of 0.13, 0.18 and 0.24 cts/cm$^2$/s for *d* = 1, 2 and 2.6 mm respectively, corresponding to a total ACD rejection efficiency of 95.8%, 94.1% and 92.2%. Reducing the distance between the ACD and the TES array the ACD can intercept and higher fraction of the trajectories passing through the main detector and therefore the geometrical rejection efficiency increases. From now on we will refer to the results obtained with *d* = 2 mm, the nominal distance in the focal plane configuration for IXO.

From the sphere model simulation the total background count rate is 0.19 cts/cm$^2$/s, slightly below the 0.2 cts/cm$^2$/s requirement, and is caused mainly by cosmic and solar protons (95% of the total background). The weight of each component is reported in Table 4. The spectra of the different components are reported in fig 4. On the total unrejected background we performed an analysis of the different particles contribution and their geometrical origin, the results of which are reported in fig 5. Since the secondaries are blocked by materials interposed external spheres do not contribute significantly to the background. Note also that about 1/5 of the residual background is due to primaries that do not cross the anticoincidence. These events are particles with skew trajectories that cross the TES array border and miss the ACD, and therefore are almost completely located in the outer array.

The total level of the unrejected background calculated with the sphere model is slightly below the requirement and it is mostly caused by secondary electrons from the inner Nb sphere and primary protons.

[3] With this energy threshold we obtained that only a 0.45% of detections in the TES array are caused by particles generated in ACD and non vetoed by the ACD itself.

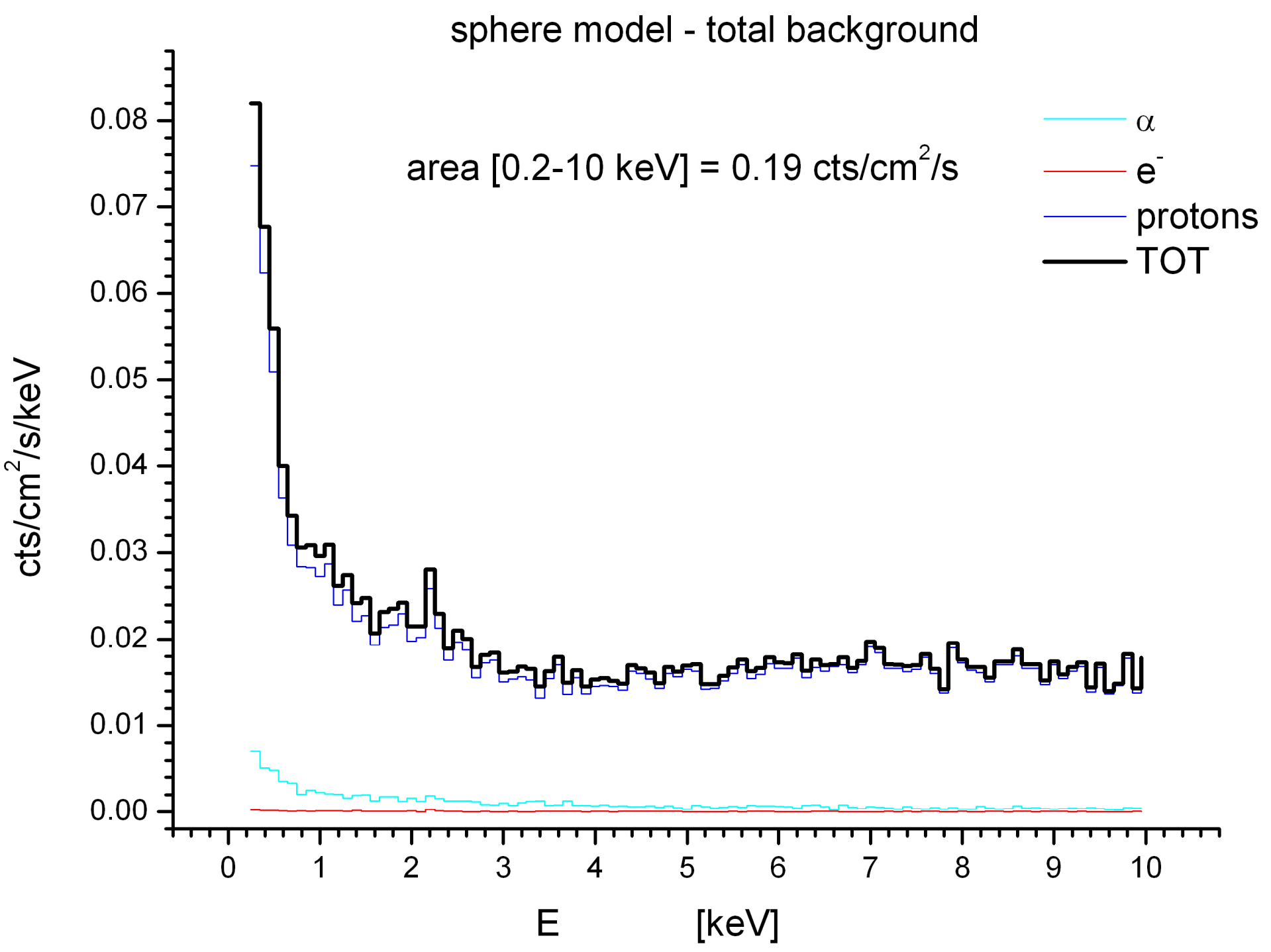

*Fig. 4. The spectrum of unvetoed events (100 eV bin), with contributions from different input components highlighted. Emission lines from nearby materials (Nb L at 2.1 keV) are visible, while the Al emission lines are blocked by the Nb coating. Fluorescence lines from Si wafer are rejected via auto-rejection, while Si line photons from ACD are vetoed by ACD itself.*

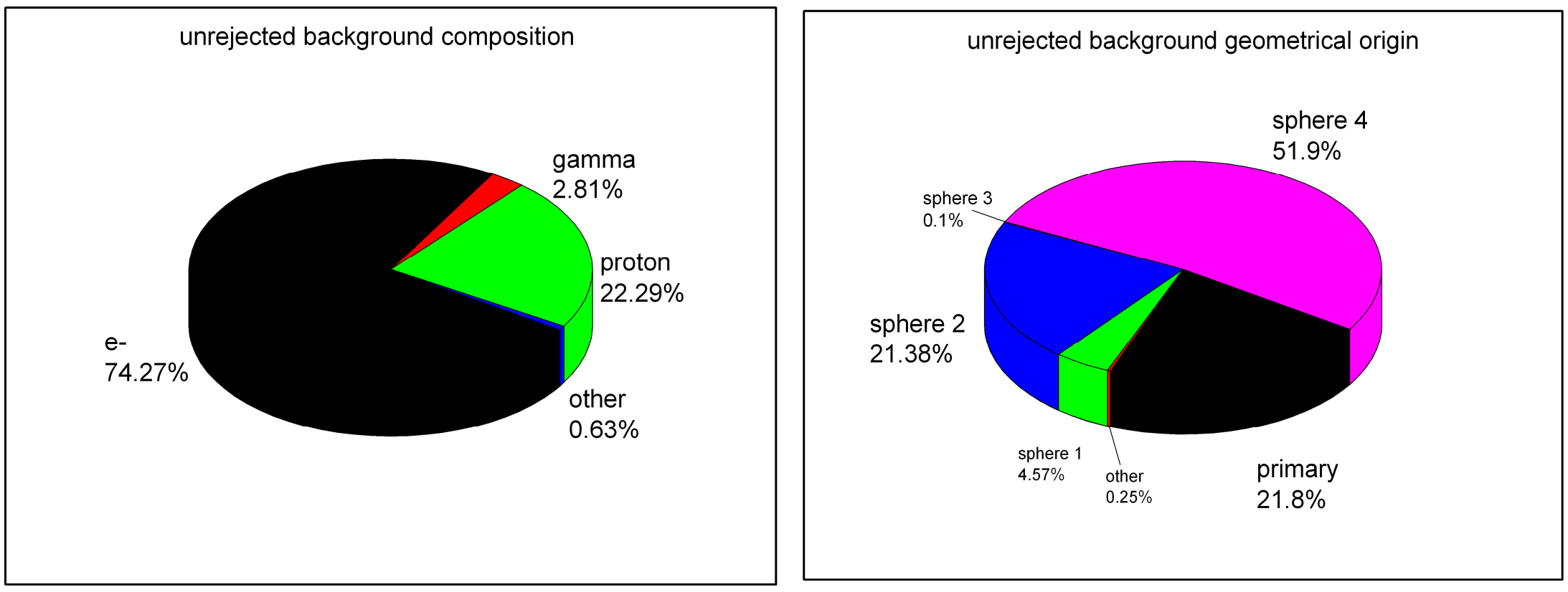

*Fig. 5. (left) – Contribution of different particles type on the unrejected background. The major contribution is given by secondary electrons, while approximately 19% is given by unrejected primary protons. (right) the geometrical origin of unrejected particles. Labels referred to fig 2.*

### *5.2. Detailed model results*

The same simulation has been run on the detailed model. The total background count rate rises to 0.31 cts/cm$^2$/s, mainly caused by particles generated by cosmic and solar protons (90% of the total background). The weight of each component is reported in Table 5 while the corresponding spectra of the different components are reported in fig 6. As for the sphere model we performed an analysis of the different particles contribution and their geometrical origin, the results of which are reported in fig 7. Most of the unrejected background (~ 80%) is due to secondary electrons. The last surface seen by the detector is the main source of background, while structures far from the detector do not contribute significantly, because the secondaries produced are blocked by the materials interposed. This result confirms that the choice to do not represent the mirrors in the outer sphere did not introduce significant differences in the background. The insertion of the TES array and ACD supports enhanced the production of secondary particles near the detector introducing a new and not negligible component of the background.

Since the dominant part of the unrejected background is induced by cosmic and solar protons, we show the spectra of different components of the proton induced background in fig 8b, 8c and 8d and the total proton induced background spectrum in figure 8a. As highlighted before the secondary electrons constitute the major part of the background, shaping the background spectra as a power law with index $\alpha \sim -0.7$ (figure 8c). The contribute of photons to the background (figure 8b) is relevant at the energy of fluorescence lines from surrounding materials i.e. Nb L line at 2.1 keV from the Nb shield and Cu K line at 8 keV from the detector supports, and completely negligible at different energies. The lines at 3.5 keV and 5.7 keV are internal lines: they result from Nb K$\alpha$ photons (~16.6 keV) impacting on the Bi of the absorber and generating L fluorescence photons (L$\alpha \sim$ 10.8 keV, L$\beta \sim$ 13 keV) that leave the absorber. The energy difference is fixed and remains inside the absorber resembling a spectral line. Even if marginal on the whole detector, the unrejected primary protons components is concentrated at high energies (see figure 8d) and in a corona composed of the outer 5 pixel of the array, where the geometrical rejection efficiency of the ACD is less efficient and the background level is some above 0.02 cts/cm$^2$/s.

Table 4. Unrejected background induced from the different primary particles: detailed model

| Primaries | Rate (cts cm$^2$ s$^{-1}$) $d$ = 2 mm |
|---|---|
| Protons | 0.28 |
| Electrons | 0.018 |
| Alpha | 9.8 x 10$^{-3}$ |
| **Total** | 0.31 |

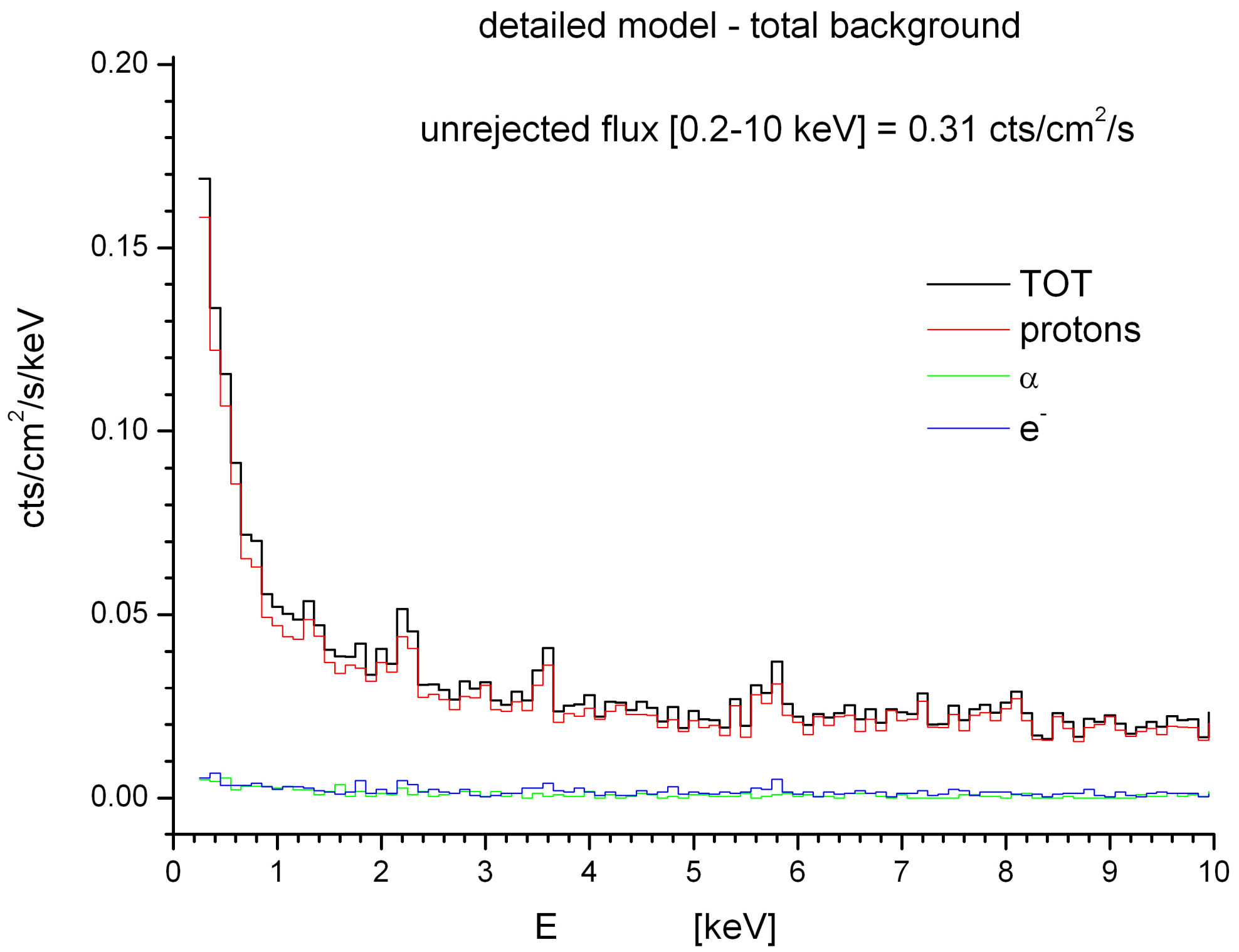

*Fig. 6. The spectrum of unvetoed events (100 eV bin) with contributions from different input components highlighted. Emission lines from nearby materials (Nb L line at 2.1 keV) are visible, while the emission lines from external materials are blocked by the Nb shield. Fluorescence lines from Si wafer are rejected via auto-rejection, while Si line photons from ACD are vetoed by ACD itself. There are two more lines visible at 3.6 and 5.8 keV, see fig 8 (2) for details.*

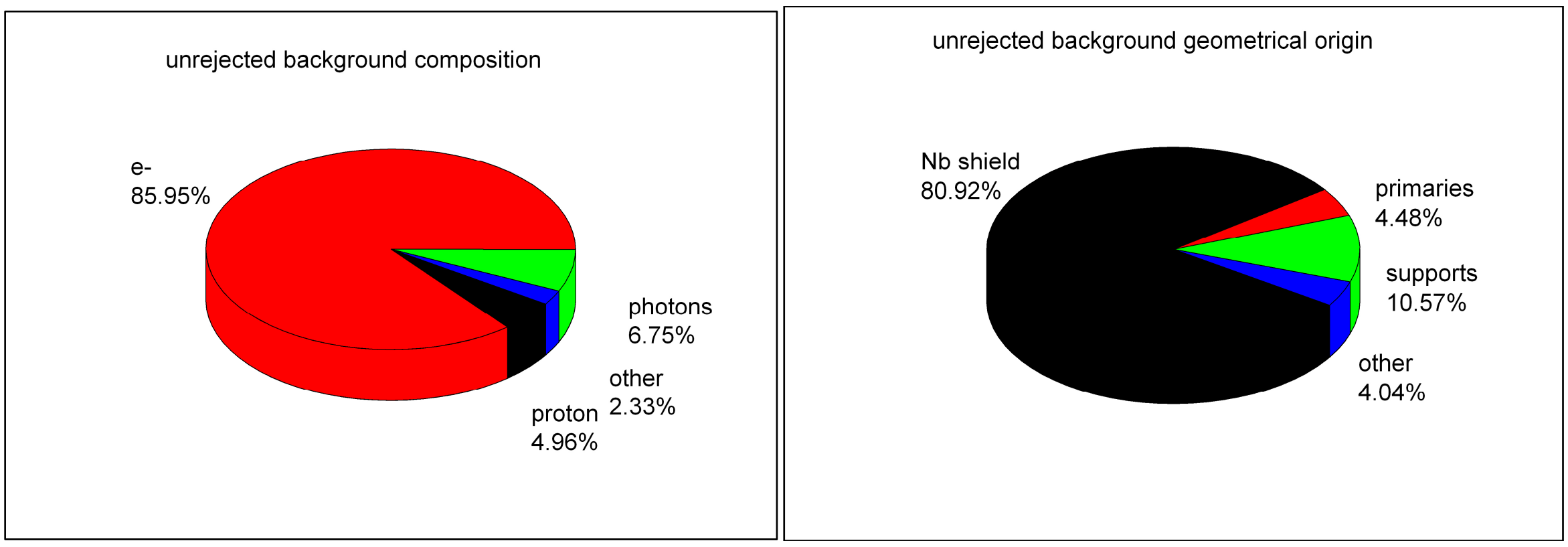

*Fig. 7. (left) – Contribution of different particles type on the unrejected background. The major contribution is given by secondary electrons, while approximately 5% is given by unrejected primary protons. (right) The geometrical origin of unrejected particles. Labels referred to fig 3.*

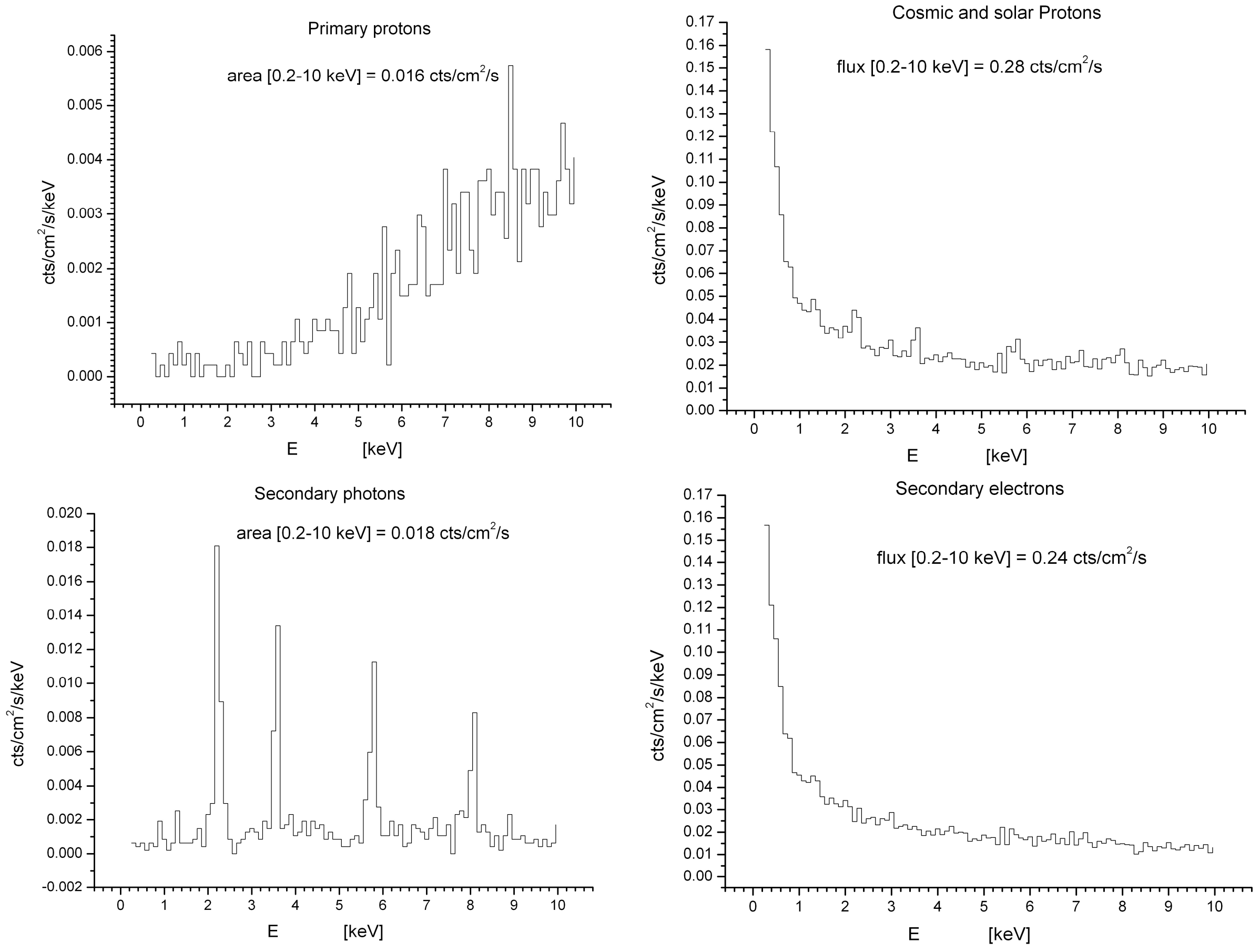

*Fig. 8. The unrejected background induced by cosmic protons:*

*a) The total background spectrum induced by CR protons*
*b) The secondary photons component, relevant at the energy of emission lines (Nb L line at 2.1 keV from the Nb shield and Cu K line at 8 keV from the detector supports), and completely marginal at different energies.*
*c) At low energies the background is caused mainly by secondary electrons generated in the Nb shield.*
*d) The unrejected primary protons contribution is concentrated at high energies*

## *5.3. Considerations on the background differences in the two geometrical models adopted*

Of particular interest is the difference in the residual background obtained by the two configurations. Due to the different geometry, and to a lesser extent to the insertion of the supports, the contribution from secondary electron (from input protons) has risen up almost by a factor 2 with respect to the level calculated with the sphere model (0.14 cts/cm$^2$/s). This enhanced contribution of electrons to the total background is due to a lower discrimination efficiency of the ACD for the secondaries produced far from the detector. In fact one should note that a secondary particle, even when fully absorbed in the TES array, can be discriminated if its primary triggers the ACD. The probability that this happens decreases with the distance, because the two paths of the primary and secondary particles can diverge significantly. In the sphere model, where the Nb coating is very close to the detector, a larger fraction of the secondaries is discriminated in this way. On the other hand, since in the detailed model the cryostat walls are more distant, there is a rise of the background of unrejected secondary particles.

## 6. Conclusions

We utilized the Geant4 toolkit to compute the background to be expected for the X-ray Microcalorimeter Spectrometer of the IXO mission in L2 point, where no experimental data are yet available for X-ray missions. We find that an active anticoincidence system is mandatory to reduce the background level on the TES array by a factor 15 to the level (~ 0.2 cts/cm$^2$/s) required to achieve the scientific performances of the instrument. The rejection efficiency increases reducing the distance of the ACD from the main detector. Low energy protons focused by the optics add only a minor contribution (0.01 cts/cm$^2$/s) to the unrejected background [14].

We used two different spacecraft mass models, a simplified and a more detailed model, and estimated the environment in L2 using the CREME96 toolkit. The comparison between background spectra obtained with the two geometrical models brought different background levels, but a similar spectral shape (see fig. 9). The background level in the detailed model is higher because the ACD efficiency in rejecting secondary particles is reduced by the increased distance of the last surface seen by the detector, as explained in Section 5.3. An accurate geometrical modeling is therefore necessary to identify the unrejected background origin and composition and to find efficient approaches to background reduction. The spectral shape is the same because, apart from the supports inserted in the complex model, the material producing the biggest fraction of the background is Nb in both cases.

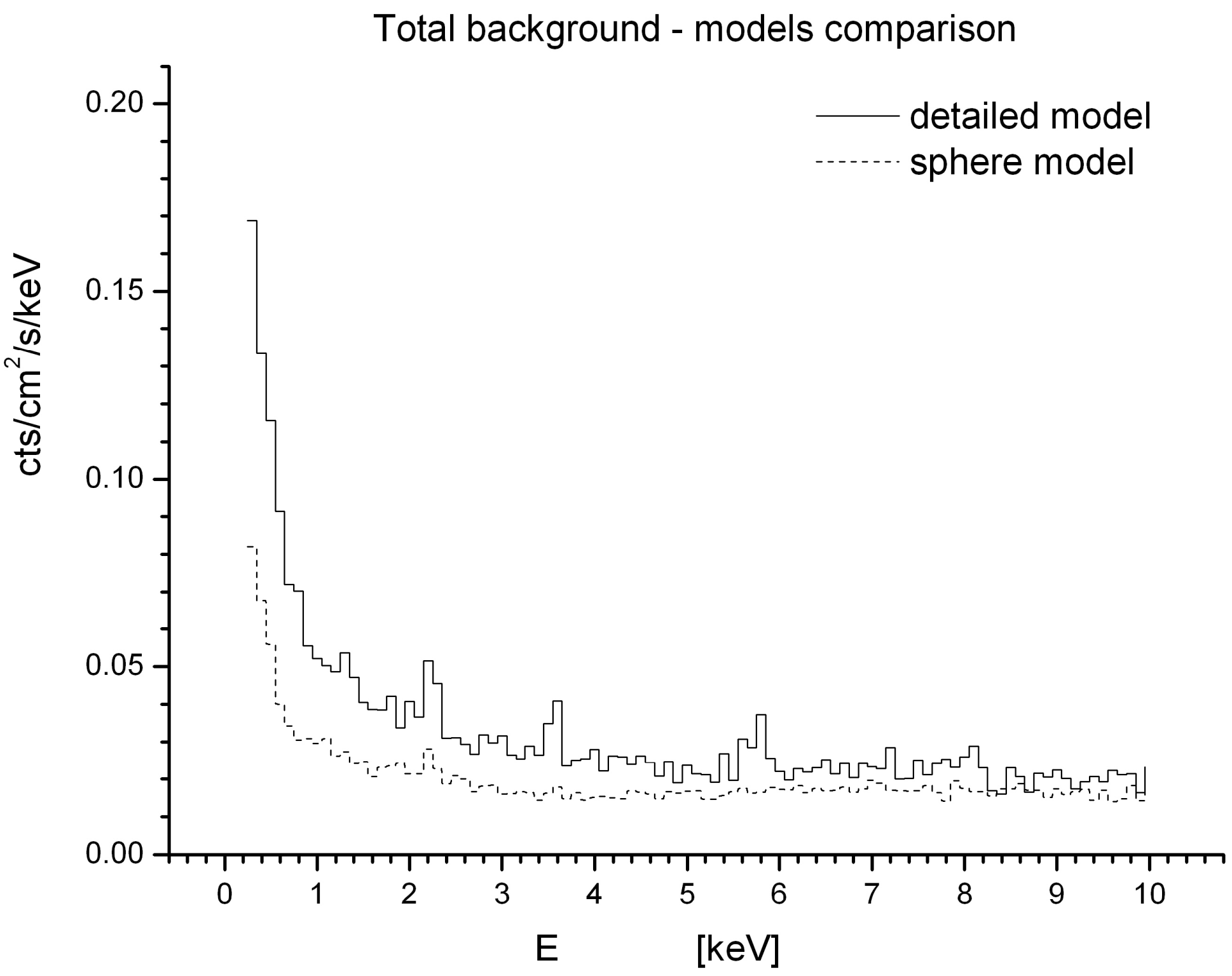

*Fig. 9. Spectral shape comparison between the two models*

The results obtained with the detailed model (0.31 cts/cm$^2$/s) were calculated using the highest fluxes levels that the IXO mission will experience during its lifetime and therefore represents the worst case scenario. They are above the IXO requirement, but we expect the final background level for ATHENA to be significantly lower than this level, thanks to a more convenient geometrical configuration and to several solutions identified with this work and discussed at the end of this section, that will allow to reduce significantly the background level to be expected on XMS. The final setup for ATHENA includes a smaller TES array, placed closer to the anticoincidence (~1 mm) that enhances the ACD rejection efficiency. The results presented here hence constitute a conservative estimate of the real background for ATHENA XMS and establish a solid base for the future simulations in the new geometrical configuration.

We showed that the background is dominated by secondary particles produced by the passage of cosmic rays through the materials surrounding the detector. In particular electrons coming from the Nb shield, the surface directly seen by the detector, turned out to constitute more than 80% of the unrejected background and this allowed to identify the need for a passive electron liner to significantly reduce the background level. A not negligible component of the background (<10%) has been introduced by the insertion of the supports, highlighting the importance of modeling the structures near the detector with great accuracy.

Note that the background is not uniformly distributed along the energy band, but is concentrated at low energies (about 50% in the 0.2-2 keV band).

With this work we have therefore identified potential areas of improvement for the background. One option, that promises to reduce the residual background, employs a thin filter very close to the array surface to stop low energy electrons: the filter will block electrons from the Nb shield, and the electrons created in the filter will be vetoed by their own primary passing through the ACD or the TES array, as explained in Section 5.3. Preliminary simulations with a 0.1 μm kapton filter (85% transmission at 0.5 keV) have predicted a background drop of ~33%. This results are promising but further investigation is needed in order to identify the most appropriate material and thickness for the filter in order to reduce the particle background without compromising photon transmission at low energies.

Providing a passive layer for electron shielding onto the innermost surfaces close to the detector can also allow to reduce the background level. Polymide films such as Kapton, or ceramics like Boron carbide seems to be the best choice, providing an high electron absorbance toward external electrons and a low electron generation rate in order to prevent the generation of new electrons towards the focal plane. Preliminary simulations have predicted a background drop of about a factor 50% with a 250 μm Kapton liner in the Nb shield [15].

## Acknowledgments

This work was supported by ASI contract I/035/10/0-10-308